\documentclass[fleqn,usenatbib]{mnras}
\usepackage{amsmath}
\usepackage{multirow}
\usepackage{graphicx}
\usepackage[authoryear]{natbib}
\usepackage{amssymb,txfonts}
\usepackage{hyperref}
\providecommand{\tabularnewline}{\\}

\newcommand{\B}{BATSE}
\newcommand{\s}{\emph{Swift}}
\newcommand{\f}{\emph{Fermi}}

\begin{document}

\title[Luminosity function of short gamma-ray bursts]{Binary neutron star merger rate via the luminosity function of short gamma-ray bursts}

\author[D. Paul]{Debdutta Paul\thanks{dbdttpl@gmail.com}\\Tata Institute of Fundamental Research, India}

\maketitle

\global\long\def\AS{\emph{AstroSat}}
\global\long\def\Ep{E_{p}}
\global\long\def\T{T_{90}}
\global\long\def\red{\chi_{{\rm red}}^{2}}
\global\long\def\Rdot{\overset{.}{R}}
\global\long\def\pyr{\rm{\,yr^{-1}}}

\begin{abstract}
The luminosity function of short Gamma Ray Bursts (GRBs) is modelled by using the available catalogue data of all short GRBs (sGRBs) detected till October, 2017. The luminosities are estimated via the `pseudo-redshifts' obtained from the `Yonetoku correlation', assuming a standard delay distribution between the cosmic star formation rate and the production rate of their progenitors. While the simple powerlaw is ruled out to high confidence, the data is fit well both by exponential cutoff powerlaw and broken powerlaw models. Using the derived parameters of these models along with conservative values in the jet opening angles seen from afterglow observations, the true rate of short GRBs are derived. Assuming a short GRB is produced from each binary neutron star merger (BNSM), the rate of gravitational wave (GW) detections from these mergers are derived for the past, present and future configurations of the GW detector networks. Stringent lower limits of $1.87 \pyr$ for the aLIGO-VIRGO, and $3.11 \pyr$ for the upcoming aLIGO-VIRGO-KAGRA-LIGO/India configurations are thus derived for the BNSM rate at $68 \%$ confidence. The BNSM rates calculated from this work and that independently inferred from the observation of the only confirmed BNSM observed till date, are shown to have a mild tension; however the scenario that all BNSMs produce sGRBs cannot be ruled out.
\end{abstract}


\section{Introduction}

Based on the observed duration and hardness distribution of a sample of gamma ray bursts (GRBs) detected by \B, \cite{Kouveliotou_et_al.-1993-ApJ} found that the sample can be divided between the so-called `long' and `short' GRBs. Although the observed event rate of GRBs vary with the GRB-detector, it is $\sim 150 \pyr$ on an average for long GRBs, but around $\sim 10 \pyr$ for short GRBs (henceforth sGRBs), almost an order of magnitude smaller. Several lines of evidence suggest that the two classes have different progenitors. Being associated with Type Ic supernovae (see \cite{Woosley_and_Bloom-2006-ARA&A} for a recent review) and also exclusively located at star forming regions within their galaxies \citep{Wainwright_et_al.-2007-ApJ, Fruchter_et_al.-2006-Nature}, long GRBs are conclusively associated with the collapse and death of massive stars \citep{MacFayden_and_Woosley-1999-ApJ, Woosley_and_MacFayden-1999-A&AS}. On the other hand, the lack of supernovae associations with sGRBs, their occurrence in older and elliptical galaxies, offset from the host galaxy, etc. (see \cite{Berger-2014-sGRB_review} for a recent review) suggest that the sGRB progenitor is the merger of compact objects giving rise to relativistic jets (\cite{Eichler_et_al.-1989-Nature}, \cite{Narayan_et_al.-1992-ApJ}; also see \cite{Nakar-2007-PhR} for a review). The recent detection of GW170817 \citep{GW170817-2017} along with its electromagnetic counterparts \citep{EM170817-2017} provided conclusive evidence of this association. Thus, the estimation of the true event rate of sGRBs can help one predict the binary neutron star merger rate, and hence their detection by gravitational wave (GW) detectors. It seems likely however that the central engine and the radiation mechanism are similar for the two classes of GRBs \citep{Ghirlanda_et_al.-2009-A&A, Calderone_et_al.-2015-MNRAS}.

The observed rate of sGRBs depends on three criterion: (1) the true event rate of GRBs as a function of intrinsic properties of the bursts, e.g. the redshift, luminosity; (2) considering GRBs are relativistic jets, the relativistic beaming of the emission reduces the observed rate from the true event rate, hence the beaming factor; and (3) the observation windows (e.g. mission time, field-of-view) and detection criteria of the GRB monitors.

Several authors have sometimes modelled different observational entities to estimate the event rate, at other times used different models for the same entity, and most obviously, different databases. For example, \cite{Guetta_and_Piran-2005-A&A}, \cite{Guetta_and_Piran-2006-A&A} and \cite{Salvaterra_et_al.-2008-MNRAS} have modelled the observed peak flux distribution of the GRBs detected by the \emph{Compton Gamma Ray Observatory (CGRO)}-\B\ \citep{Fishman_et_al.-1989-BAAS}, while \cite{Hopman_et_al.-2006-ApJ}, \cite{Guetta_and_Stella-2009-A&A}, \cite{Dietz-2011-A&A} and \cite{Petrillo_et_al.-2013-ApJ} have modelled the distribution of observed redshifts. The latter approach may be severely affected by the detection of only a handful of bursts with redshifts measurements, with the measurement of the redshifts itself being biased towards smaller values via the identification of host galaxies. Such selection biases have been studied and alternative approaches proposed by \cite{D'Avanzo_et_al.-2014-MNRAS}. On the other hand, \cite{Virgili_et_al.-2011-ApJ} have attempted to fit the peak-flux distribution of both \B\, and \s\, \citep{Barthelmy_et_al.-2005, Gehrels_et_al.-2004-ApJ} bursts, as well as the observed redshift distribution, while \cite{Wanderman_and_Piran-2015-MNRAS} have used the peak-flux distribution of \B, \s\, and \f -GBM \citep{FermiGBM-2009-ApJ} bursts along with the redshifts distribution. Subsequently, different authors have placed different constraints on the true event rate. While \cite{Guetta_and_Piran-2005-A&A} reported the rate in the local universe, $\Rdot(0)$, to be $0.1$-$ 0.8 \, \rm{ yr^{-1} Gpc^{-3} } $, \cite{Guetta_and_Piran-2006-A&A} extended it to be $8$-$ 30 \, \rm{ yr^{-1} Gpc^{-3} } $ with the addition of \s\, and HETE II bursts, and \cite{Coward_et_al.-2012-MNRAS} at $5$-$ 13 \, \rm{ yr^{-1} Gpc^{-3} } $ from \s\, bursts alone. On the other hand, \cite{Salvaterra_et_al.-2008-MNRAS}, \cite{Virgili_et_al.-2011-ApJ} and \cite{Wanderman_and_Piran-2015-MNRAS} have claimed that progenitors other than compact object mergers are required to model the detected distributions.

The most straightforward way of modelling the event rate is by modelling the luminosity distribution directly, in that the fundamental parameters that create the observed rate distributions for all observed parameters are the redshift ($z$) and luminosity ($L$). The number of GRBs detected by an instrument in the redshift range $z_1$ to $z_2$ and luminosity range $L_1$ to $L_2$ can be written as:

\begin{equation}
N(L_{1},L_{2};z_{1},z_{2}) = T \, \dfrac{\Delta\Omega}{4\pi} \,
\intop_{z_{1}}^{z_{2}} \dfrac{\Rdot(z)}{1+z} dV
\intop_{\rm{max}[L_1,\, L_c]}^{L_2} \Phi_z(L) dL,
\label{eq:definition_of_phi}
\end{equation} where $T$ is the duration of operation of the instrument; $\Delta \Omega$ its field-of-view; $L_{c}(z)$ denotes its lower-cutoff in the detectable luminosity set by its flux-sensitivity limit $ P_{ \rm{lim} } , $ given by  $L_c (z) = P_{ \rm{lim} } \, 4\pi d_L^2(z) $  (see Fig. \ref{fig:L-z}); the rate of GRBs beamed towards the observer from an infinitesimal co-moving volume $dV$ is given by $\Rdot(z)\frac{dV}{1+z},$ the factor $(1+z)^{-1}$ taking care of the cosmological time dilation. The probability density function $\Phi_z(L),$ formally called the `luminosity function' (henceforth LF), has the unit of $\rm{ ( erg \, s^{-1} )^{-1} }, $ the subscript referring to an implicit dependence on the redshift. Now, $\Rdot\left(z\right)$ can be written as
\begin{equation}
\Rdot\left(z\right) = f_{\rm{B}} C\, \Psi(z),
\label{eq:R_dot}
\end{equation} where $\Psi(z)$ gives the mass of the GRB progenitors available per unit time per unit volume (in units of $ \rm{ M_{\odot} yr^{-1} Gpc^{-3} } $), $C$ gives the efficiency of their production per unit available mass (in units of ${\rm M_{\odot}^{-1}}$), and $f_{\rm{B}}$ is the beaming factor of the relativistic jets responsible for the burst. $\Psi(z)$ is extensively discussed in Section \ref{sec:The luminosity function}. As far as the modelling of the LF is concerned, the parameters $C$ and $f_{\rm{B}}$ are degenerate; an extensive discussion on this is deferred to Section \ref{sec:Event rate}.

Direct modelling of the LF via the luminosity distribution suffers from the fact there are too few GRBs with observed redshifts and hence estimated luminosities, moreover the sample can suffer from heavy selection bias for the redshift measurement. Although \cite{D'Avanzo_et_al.-2014-MNRAS} suggested a method of eliminating the selection bias by limiting to a `flux-complete' sample, the number of bursts thus obtained is too low to make direct modelling of the LF meaningful. To get around this problem, \cite{Yonetoku_et_al.-2004-ApJ} originally proposed a method for long GRBs. Whereas \cite{Amati_et_al.-2002-A&A} had found a correlation between the total isotropic energy in the source frame ($E_{iso}$) and the spectral energy peak of the Band function \citep{Band_et_al.-1993-ApJ} $E_p$, they found that the correlation gets tighter between the observed peak luminosity ($L_{iso}$) and the source frame-corrected $ E_{p,0} = E_p (1+z) $. Assuming that this correlation is followed by all long GRBs, they estimated `pseudo-redshifts' of $689$ long GRBs in the BATSE sample and were thus able to study the luminosity distribution of these GRBs directly. Recently, this method was extended by \cite{Yonetoku_et_al.-2014-ApJ} on sGRBs, using the so-called `Yonetoku correlation' found by \cite{Tsutsui_et_al.-2013-MNRAS} for eight sGRBs. They used $72$ \B\, sGRBs whose spectra were modelled by the Band function, concluding that the LF is consistent with a simple powerlaw with an index of unity, and $\Rdot(0)$ in the range $ 0.24$-$0.94 \; \rm{ yr^{-1} Gpc^{-3} } $.

\cite{Ghirlanda_et_al.-2016-A&A} did an extensive study of the distributions of four observed parameters of \s\, and \f\, bursts, namely the peak flux, fluence, observer frame duration and the observer frame peak energy, and also the distributions of redshift, isotropic energy and isotropic luminosity of a `flux-complete' sample of \s\, bursts presented by \cite{D'Avanzo_et_al.-2014-MNRAS}. In doing so, they assumed the validity of the Yonetoku as well as the Amati correlations, whose parameters were included in the model. Contradictory to \cite{Yonetoku_et_al.-2014-ApJ}, they concluded that the LF is inconsistent with a simple powerlaw function, and fitted a broken powerlaw with a constant break luminosity, combined with different distributions of the formation rate of the sGRB progenitors. They reported $ \Rdot(0) \simeq 0.13$-$0.24 \; \rm{ yr^{-1} Gpc^{-3} } $ and $ \Rdot(0) \simeq 0.65$-$1.1 \; \rm{ yr^{-1} Gpc^{-3} } $ for the two fitted models.

In this work, I have applied the method followed by \cite{Yonetoku_et_al.-2014-ApJ} to model the luminosity distribution of the full sample of sGRBs detected by \B, \f\, and \s\, till October, 2017. This is made possible by using a simplification proposed for long GRBs by \cite{Paul-2018-MNRAS}: instead of modelling the spectra of each individual GRB accurately, they are statistically sampled from the true distribution as observed for \f -GRBs, utilizing the wideband information available for \f -GBM. I have then used the fitted models to calculate the true event rate of the sGRBs, and assuming that they are produced from binary neutron star (henceforth NS) mergers, deduced the rate of electromagnetic counterparts of gravitational wave events to which the GW detectors are sensitive in their different observing phases \citep{Abbott_et_al.-2016-review}. This work simplifies the understanding of the sGRB production scenario significantly over previous works who carry out general numerical studies of all the parameters in the problem. In assuming inputs from the star formation history of the universe and population synthesis models, and  resorting to direct observations wherever applicable (e.g. the Yonetoku correlation data and the \f\, spectral parameter observations), it considerably simplifies the numerical framework and demonstrates that robust statements about the physical scenario can be made nonetheless.

This paper is organized as follows. The validity of the Yonetoku correlation is investigated in Section \ref{subsec:Yonetoku correlation}, the generation of the luminosity of all sGRBs is described in Section \ref{subsec:The L-z data generation}, the modelling of the LF is detailed in Section \ref{subsec:Modelling the LF}, the local GRB rate is inferred from the models in Section \ref{subsec:The local GRB rate}, and predictions are made for \AS -CZTI in Section \ref{subsec:CZTI predictions}. In Section \ref{sec:Event rate}, the true sGRB rate is derived via the derived models, and extrapolated to derive the BNSM rate, and in Section \ref{sec:Conclusions}, concluding remarks are presented. Throughout this paper, a standard $\Lambda$-CDM (cold dark matter) cosmology with $ H_0 = 72 \, \rm{km \, s^{-1} Mpc^{-1}} ,$ $ \Omega_m = 0.27 $ and $ \Omega_{\Lambda} = 0.73 $ has been assumed. All the catalogue data, scripts used and important databases generated are publicly available at \url{https://github.com/DebduttaPaul/luminosity_function_of_sGRBs}.

\begin{table*}
\caption{The catalogue of 15 short GRBs with well-measured redshift and spectral parameters, defined as $ \T \leq 2.0 $ s. The spectral peak given here refers to that in the source frame. Data taken from:
[1] \citet{Tsutsui_et_al.-2013-MNRAS}, [2] \citet{Paul-2018-MNRAS}, [3] \citet{D'Avanzo_et_al.-2014-MNRAS} and references therein.
\label{tab:short_catalogue}}
\begin{center}
\begin{tabular}{|c|c|c|c|c|c|}
\hline
GRB name & $T_{\rm{90}}$ & $z$ & $E_{p,0}$ & $L_p$ & reference \tabularnewline
 & [s]&  & [keV] & [$ 10^{50}\, \rm{erg \, s^{-1}} $] & \tabularnewline
\hline
\hline
040924 & $1.51$ & $0.86$ & ${124.55}^{+11.15}_{-11.15}$ & $228^{+25}_{-24}$ & 1 \tabularnewline
\hline
050709 & $0.70$ & $0.16$ & ${97.32}^{+7.76}_{-0.58}$ & $7.51^{+0.76}_{-0.81}$ & 1 \tabularnewline
\hline
051221A & $1.41$ & $0.55$ & ${621.69}^{+87.42}_{-67.69}$ & $277^{+29}_{-29}$ & 1 \tabularnewline
\hline
061006 & $0.50$ & $0.44$ & ${954.63}^{+198.39}_{-125.86}$ & $206^{+15}_{-31}$ & 1 \tabularnewline
\hline
070714B & $2.00$ & $0.92$ & ${2150.40}^{+910.39}_{-443.52}$ & $656^{+79}_{-136}$ & 1 \tabularnewline
\hline
080905A & $0.96$ & $0.12$ & ${759.30}^{+308.28}_{-308.28}$ & $1.02^{+1.02}_{-1.02}$ & 2 \tabularnewline
\hline
090510 & $0.30$ & $0.90$ & ${8679.58}^{+947.69}_{-947.69}$ & $10400^{+2400}_{-1400}$ & 1 \tabularnewline
\hline
100117A & $0.31$ & $0.92$ & ${936.96}^{+297.6}_{-297.6}$ & $189^{+21}_{-35}$ & 1 \tabularnewline
\hline
100206 & $0.13$ & $0.41$ & ${638.98}^{+131.21}_{-131.21}$ & $99.8^{+115.0}_{-32.5}$ & 1 \tabularnewline
\hline
100625A & $0.32$ & $0.45$ & ${701.32}^{+114.71}_{-114.71}$ & $34^{+1}_{-1}$ & 3 \tabularnewline
\hline
100816A & $2.00$ & $0.81$ & ${235.36}^{+15.74}_{-15.74}$ & $96.9^{+19.5}_{-12.8}$ & 1 \tabularnewline
\hline
101219A & $0.60$ & $0.72$ & ${841.82}^{+107.56}_{-82.50}$ & $156^{+24}_{-23}$ & 1 \tabularnewline
\hline
111117A & $0.46$ & $1.30$ & ${966.00}^{+322.00}_{-322.00}$ & $404^{+128}_{-128}$ & 3 \tabularnewline
\hline
130603B & $0.09$ & $0.36$ & ${894.96}^{+135.60}_{-135.60}$ & $435^{+87}_{-87}$ & 3 \tabularnewline
\hline
131004A & $1.15$ & $0.72$ & ${247.22}^{+153.72}_{-153.72}$ & $23.73^{+23.73}_{-23.73}$ & 2 \tabularnewline
\hline
\end{tabular}
\end{center}
\end{table*}

\begin{figure*}
\centering{}
\includegraphics[scale=0.5]{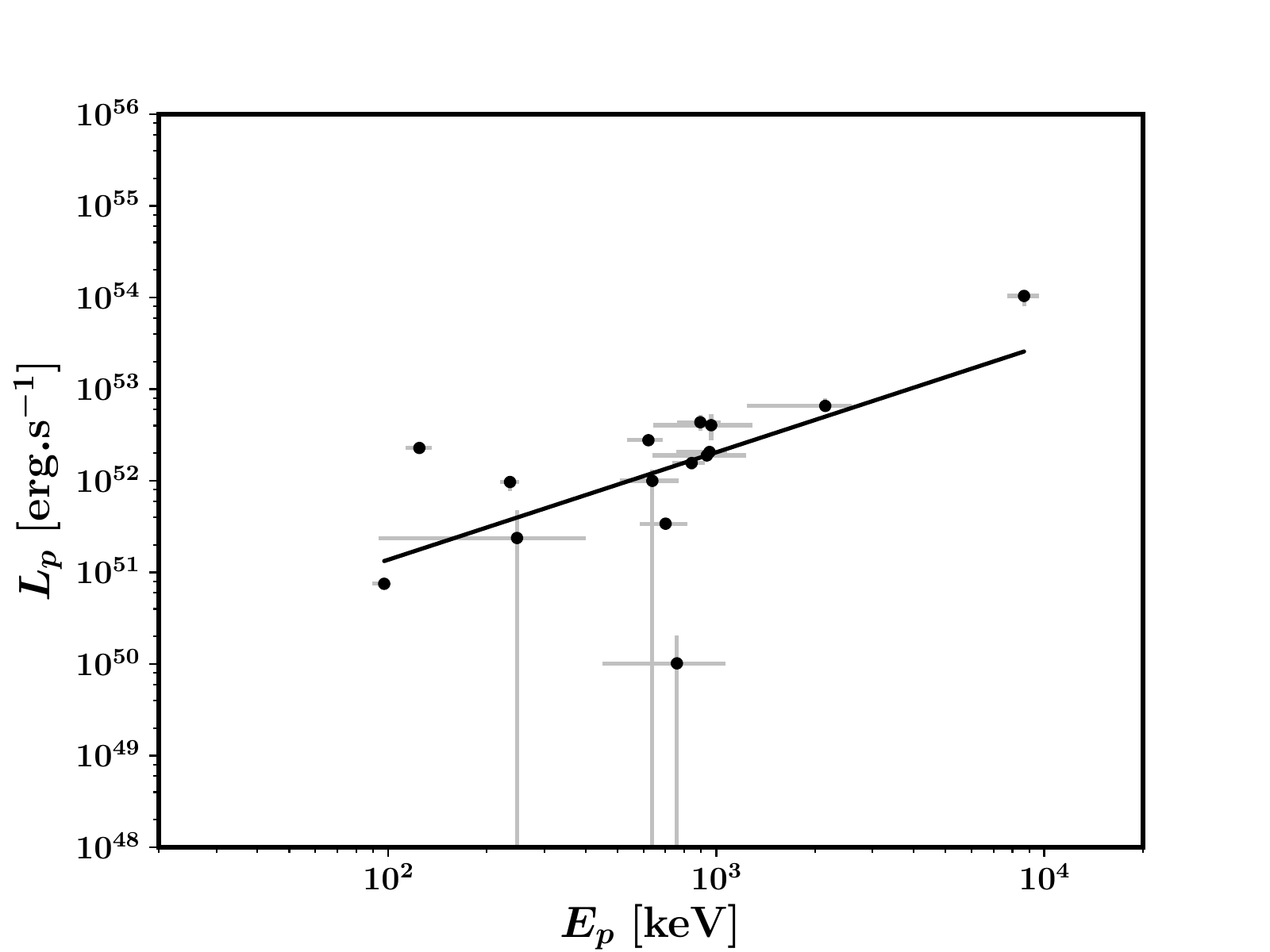}
\caption{The Yonetoku correlation as seen from the data of 15 short GRBs with spectral parameters as well as redshift measurement, given in Table \ref{tab:short_catalogue}. $ A = 2.04 \pm 0.22 $ and $ \eta = 1.17 \pm 0.18 $ corresponding to Equation \ref{eq:Yonetoku_correlation} shows the best fit as the black solid line. 
\label{fig:Yonetoku_correlation}}
\end{figure*}

\section{The luminosity function}
\label{sec:The luminosity function}

\subsection{The Yonetoku correlation}
\label{subsec:Yonetoku correlation}

The validity of the Yonetoku correlation is first tested by combining all data from existing literature. The short GRBs are defined as $\T < 2.0$ seconds, instead of using the duration in the rest frame, used by \cite{Tsutsui_et_al.-2013-MNRAS}. This is to be consistent with the general convention followed in the rest of the work, where the full sample without the redshift information are used, making it impossible to classify bursts using only the source-frame criterion. Using this criterion, we find 15 GRBs in the literature, given in Table \ref{tab:short_catalogue} and plotted in Fig. \ref{fig:Yonetoku_correlation}. Although the number of sources is small and there are at least three outliers, the correlation is found to be significant: a linear correlation coefficient of $0.98$ is retrieved, and the hypothesis that it is generated from a random distribution is discarded (a probability of $ 8.9 \times 10^{-11} $). This also justifies that the effect of outliers on the correlation is not significant, hence possible contamination of the sample by long GRBs, or the effect of missing out a few short GRBs with longer durations in the observer's frame \citep{Zhang_et_al.-2009-ApJ} does not have any effect on the rest of the work.

The best fit to the linear correlation is given by
\begin{equation}
\dfrac{L_{p}}{10^{52}{\rm \, erg \, s^{-1}}} = A \left[\dfrac{E_{p,0}}{{\rm MeV}}\right]^{\eta},
\label{eq:Yonetoku_correlation}
\end{equation} with $ A = 2.04 \pm 0.22 $ and $ \eta = 1.17 \pm 0.18. $  The parameters obtained by \cite{Tsutsui_et_al.-2013-MNRAS}, with the presently-defined normalization, are given by $ A = 2.93^{+0.57}_{-0.48} $ and $ \eta = 1.59 \pm 0.11 $. The results are thus not significantly different, as expected from the fact that the current database includes and extends their dataset.

\subsection{Generating the luminosity data}
\label{subsec:The L-z data generation}

To model the LF of sGRBs, their luminosities are required for a large sample. In this work, the Yonetoku method of estimating luminosities via the pseudo-redshifts from the Yonetoku correlation is extended to include all sGRBs available in the catalogues of \emph{CGRO}-\B\, \citep{Fishman_et_al.-1989-BAAS, BATSE_catalogue--1997}\footnote{\href{https://heasarc.gsfc.nasa.gov/W3Browse/all/batsegrb.html}{https://heasarc.gsfc.nasa.gov/W3Browse/all/batsegrb.html}}, \s -BAT \citep{Gehrels_et_al.-2004-ApJ, Barthelmy_et_al.-2005}\footnote{\href{https://swift.gsfc.nasa.gov/archive/grb_table/}{https://swift.gsfc.nasa.gov/archive/grb\_{}table/}}, and 
\f -GBM \citep{FermiGBM-2009-ApJ, Fermi_catalgoue--2016-ApJS}\footnote{\href{https://heasarc.gsfc.nasa.gov/W3Browse/fermi/fermigbrst.html}{https://heasarc.gsfc.nasa.gov/W3Browse/fermi/fermigbrst.html}}. As before, the distinction between the short and long bursts is drawn at $ \lessgtr 2 $ seconds, and a total of $757$ GRBs are thus available up to GRB171025913 (\f\, nomenclature).

\begin{table}
\caption{The number of GRBs for which pseudo-redshifts are estimated, for each mission. For both \B\, and \s, the spectral parameters are not available in the catalogues. The \f\, catalogue however contains GRBs with both Band function parameters estimated, and otherwise; the classification of \f -bursts is explained in Section \ref{subsec:The L-z data generation} and are plotted separately in Fig. \ref{fig:L-z}.
\label{tab:pseudo_numbers}}
\begin{center}
\begin{tabular}{|c|c|c|}
\hline 
mission & spectral parameters available & number \tabularnewline
\hline 
\hline 
\B &  no & $468$ \tabularnewline
\hline 
\multirow{2}{*}{\f} & yes, Type I & $188$ \tabularnewline
\cline{2-3} 
 & no, Type II & $21$ \tabularnewline
\hline 
\s &  no & $59$ \tabularnewline
\hline
\hline
TOTAL & & $736$ \tabularnewline
\hline
\end{tabular}
\end{center}
\end{table}

\begin{table}
\caption{The number of GRBs available for each mission. See Table \ref{tab:pseudo_numbers} for the classification of GRBs with pseudo redshifts. The nomenclature here also corresponds to the k-correction used for the calculation of the luminosities (see Equations \ref{eq:definition_of_k--Fermi}, \ref{eq:definition_of_k---Swift}) as well as the selection thresholds, $L_{c}(z)$ (see Equation \ref{eq:definition_of_phi} and Fig. \ref{fig:L-z}).
\label{tab:GRBs_used_for_modelling_LF}}
\begin{center}
\begin{tabular}{|c|c|c|}
\hline
mission & redshift & number \tabularnewline
\hline
\hline
\B &  pseudo & $468$ \tabularnewline
\hline
\multirow{2}{*}{\f} & pseudo & $209$ \tabularnewline
\cline{2-3} 
 & measured & $2$ \tabularnewline
\hline
\multirow{2}{*}{\s} & pseudo & $59$ \tabularnewline
\cline{2-3}
 & measured & $19$ \tabularnewline
\hline
\hline
TOTAL & & $757$ \tabularnewline
\hline
\end{tabular}
\end{center}
\end{table}

\begin{figure*}
\centering{}
\includegraphics[scale=0.50]{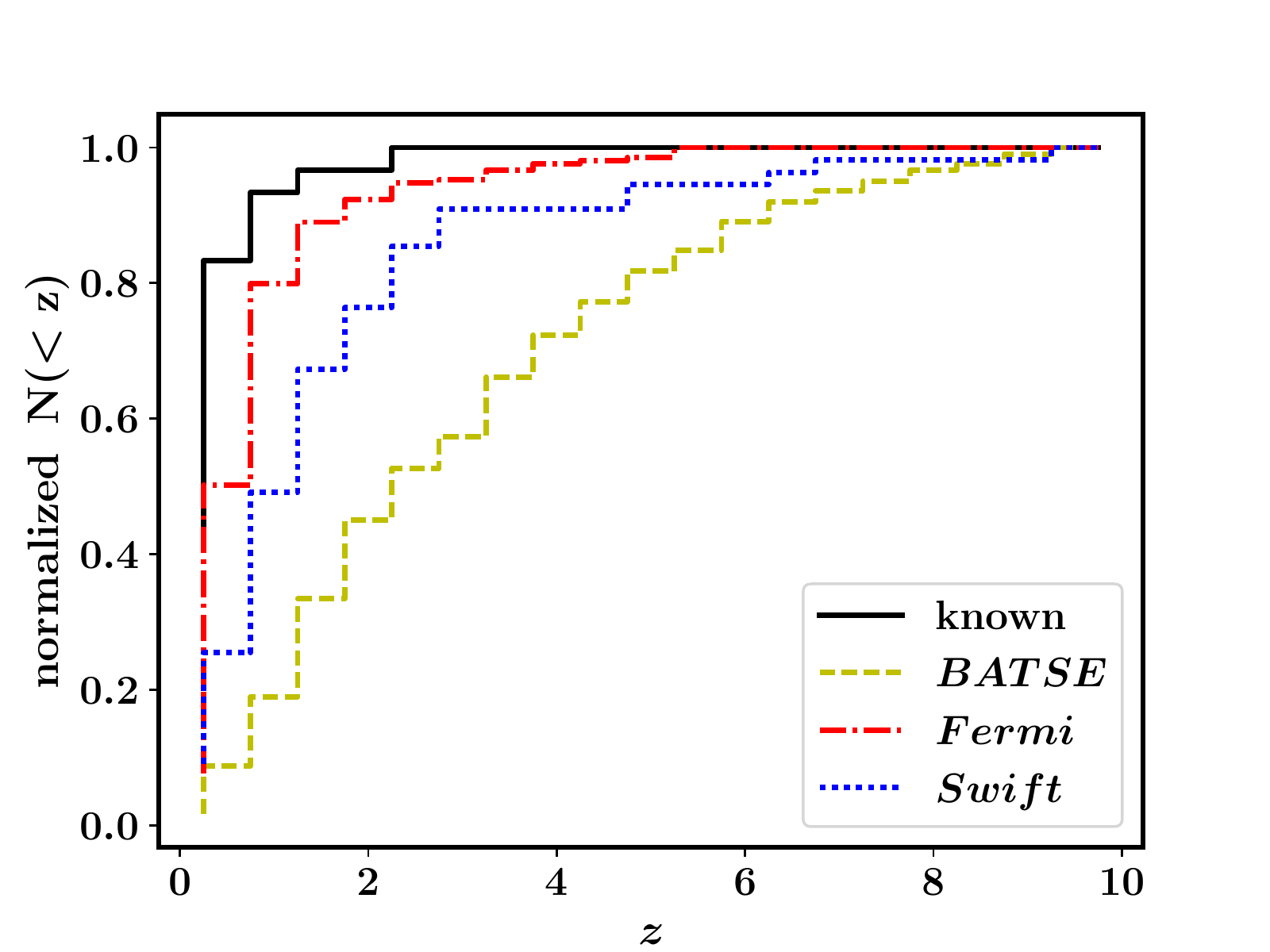}
\includegraphics[scale=0.50]{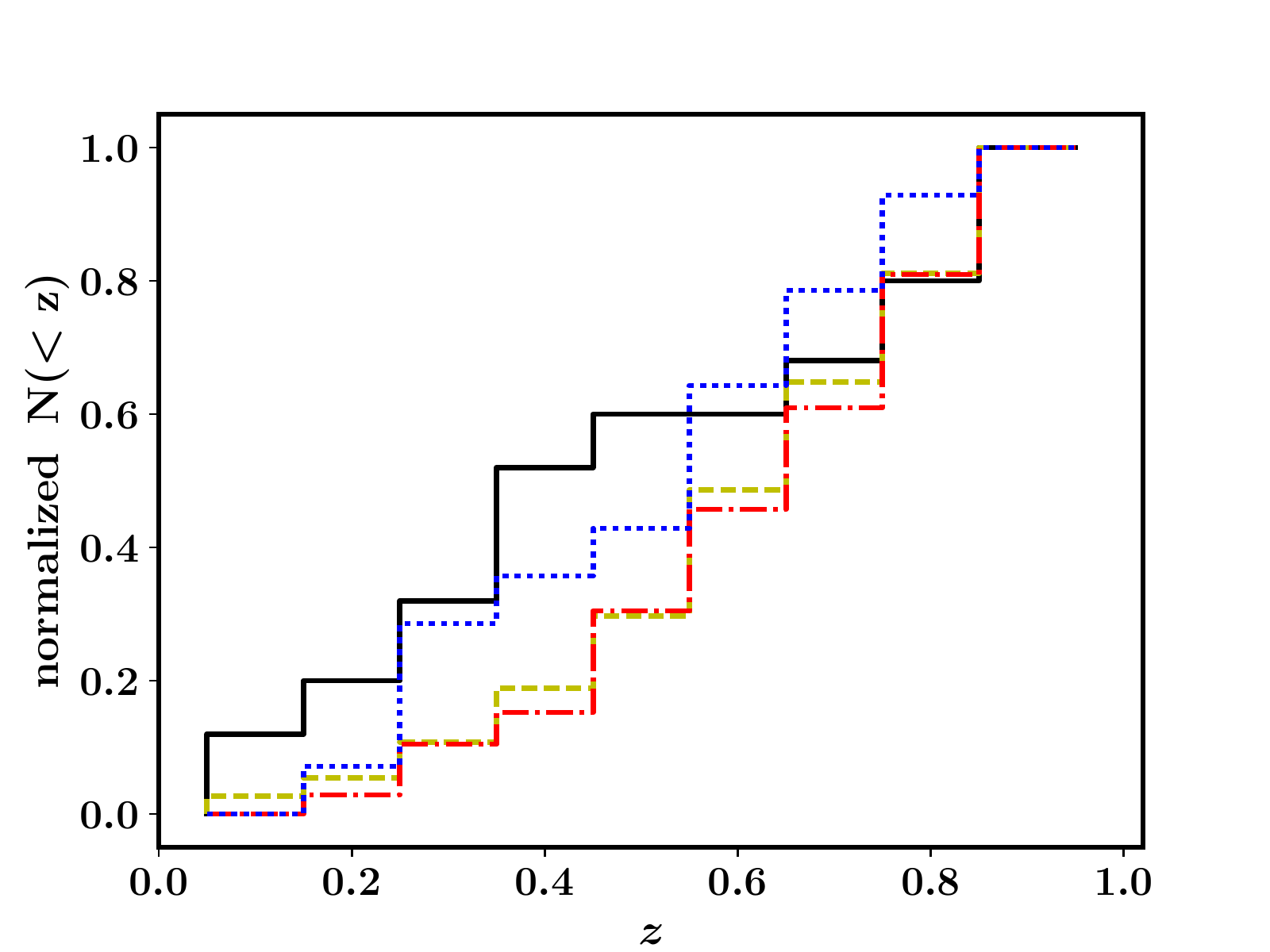}\caption{The cumulative distribution of redshifts. The distribution from the $30$ GRBs with known redshifts, including the $15$ given in Table \ref{tab:short_catalogue}, is plotted in black. The pseudo redshifts derived for \B, \f\, and \s\, GRBs are shown in yellow (dashed),  red (dot-dashed) and blue (dotted) lines respectively. \emph{Left}: The full range used for modelling the luminosity function. The 2-sample KS test rules out the hypothesis that the pseudo redshifts are derived from the same distribution as that of the known redshifts. However, the number of GRBs with known redshifts being very small ($30$), this may be due to the instrumental selection effect in redshift measurement. \emph{Right}: The same distribution truncated at a redshift of $1.0,$ below which $25$ of the GRBs with measured redshifts are located. The KS test cannot rule out that all the curves are drawn from the same distribution up to at least this redshift, upto a high degree of confidence. (A coloured version of this figure is available in the online journal.)
\label{fig:z-distribution}}
\end{figure*}

\begin{figure*}
\centering{}
\includegraphics[scale=0.50]{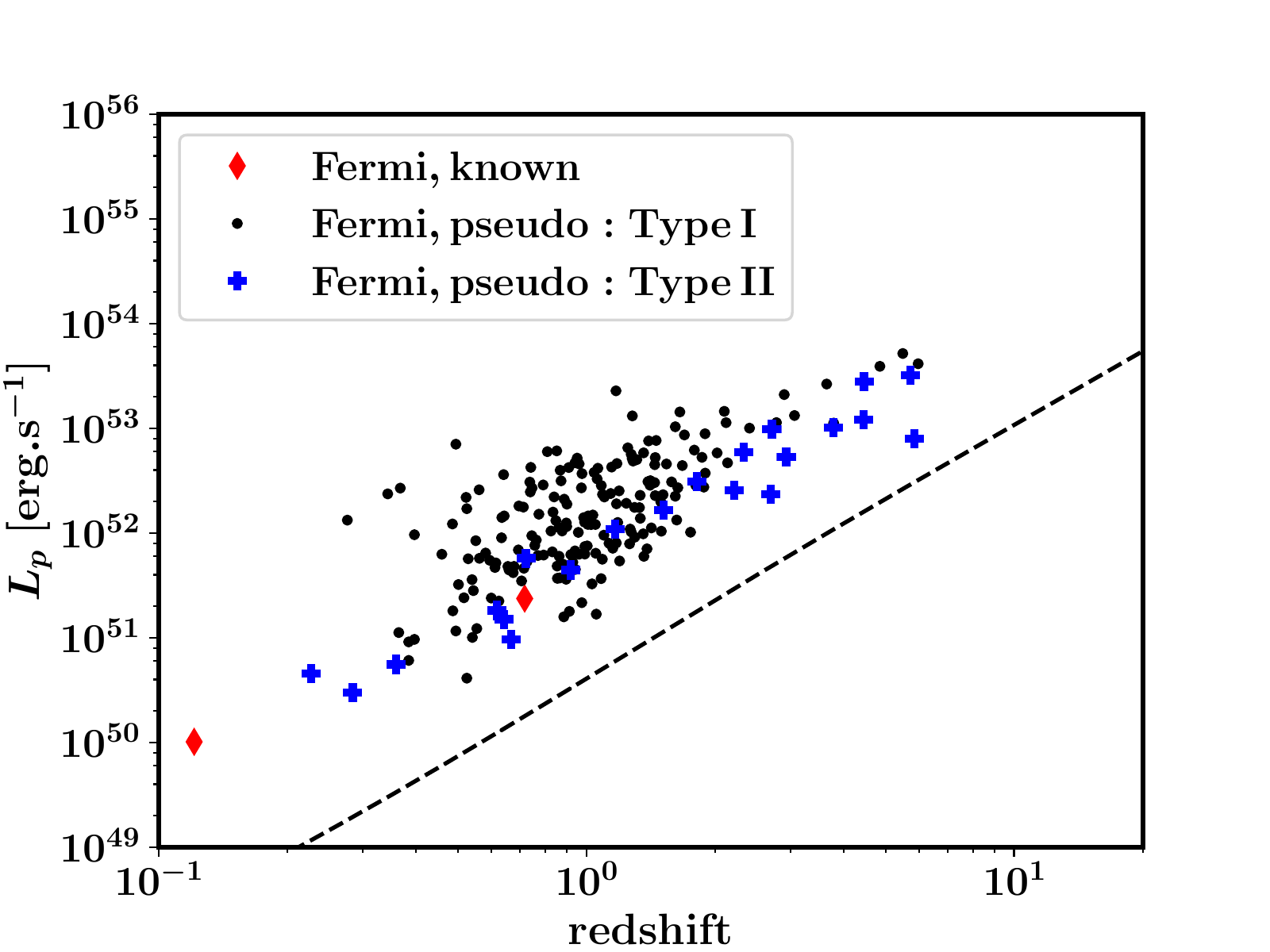}
\includegraphics[scale=0.50]{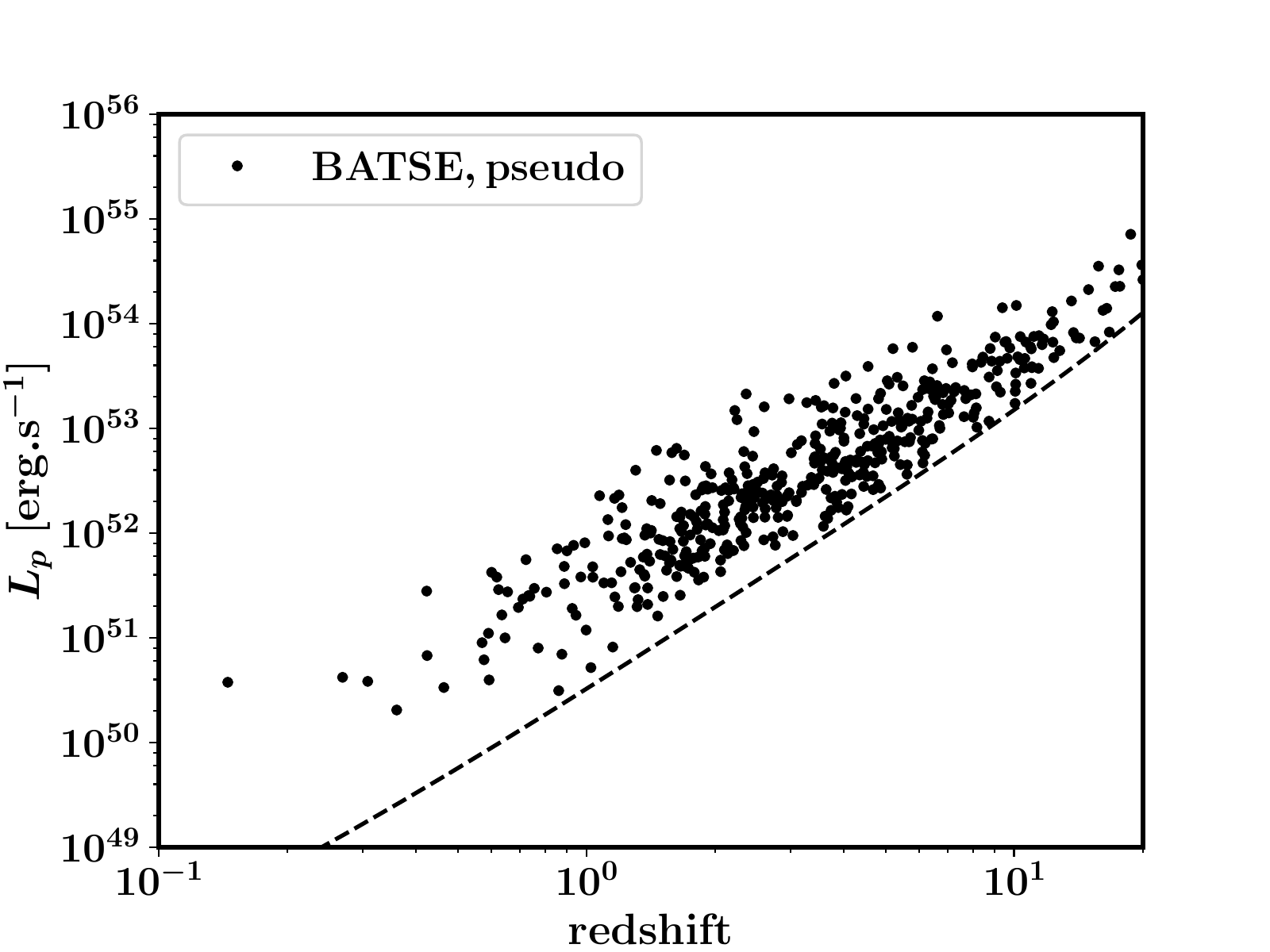}
\caption{The $L$-$z$ distributions. The dashed curves give the instrumental sensitivity limits, $L_c(z)$ for the respective instruments, see text below Equation (\ref{eq:definition_of_phi}). \emph{Left}: For \f; in red (diamond) are the two GRBs with redshift measurement from \s, in black (dot) and blue (plus) are those with pseudo redshifts measured from the Yonetoku correlation, with (Type I, black-dot) and without (Type II, blue-plus) spectral parameters available in the \f\, catalogue (see Section \ref{subsec:The L-z data generation} for the classification and Table \ref{tab:pseudo_numbers} for the corresponding numbers). \emph{Right}: Pseudo-redshifts estimated for all \B\, GRBs. (A coloured version of this figure is available in the online journal.)
\label{fig:L-z}}
\end{figure*}

Since the operational time of \B\, and the later missions are mutually exclusive, there is no \B\, GRB that is coincident with \s\, and \f. However, the latter two missions do have a small but significant number of coincidence, see \cite{Paul-2018-MNRAS} (their Section 2.1 and Fig. 2), and the same method is followed for identifying them. All bursts which are detected by both the instruments but do not have redshift measurements, are treated as \f\, GRBs and included in the \f\, dataset for the modelling, i.e. the corresponding $k(z)$ (see below) and $L_{c}(z)$ are used. For the exclusively \f -bursts, the ones with spectral parameters available in the catalogue are referred to as Type I, while the ones without them as Type II. For the latter, pseudo redshifts are generated similar to that of all other \B\, and \s\, bursts, see below. The number of bursts thus available for generating pseudo redshifts are given in Table \ref{tab:pseudo_numbers}. The total number of \s\, bursts with available redshifts, with or without spectral parameters and including those in Table \ref{tab:short_catalogue}, is $30$. Since the modelling for each mission needs to be carried out separately due to the difference in $L_{c}(z)$ (see Equation \ref{eq:definition_of_phi}), the total number available for each mission for this purpose are given separately in Table \ref{tab:GRBs_used_for_modelling_LF}.

\cite{Bromberg_et_al.-2013-ApJ} has pointed out that $ \lessgtr 0.8 $ seconds is a better classifier for \s\, GRBs, and \cite{Wanderman_and_Piran-2015-MNRAS} has supported this claim from their independent study. To estimate the effect of using this classification scheme for \s\, GRBs used in this work, I have carried out the generation of the pseudo redshifts and estimated the corresponding luminosity distribution (see below) for both the \s\, and \f\, GRBs, similar to the data plotted in Fig. \ref{fig:data-vs-model}. Executing the 2-sample KS-test on the two luminosity datasets thus generated for each of \s\, and \f\, separately, it is observed that a probability of them being drawn from the same sample is very close to unity (upto nine places after the decimal) in both cases. Hence, the different classification scheme has no bearing on the modelling of the LF and the conclusions about the event rate of sGRBs and BNSMs. This is an advantage of using a method that is only reliable in the statistical sense, as discussed below.

The peak luminosity needs to be corrected for the spectral k-correction factor as follows:

\begin{equation}
L_{p} = P \; 4\pi d_{L}(z)^{2}\times k(z;\,{\rm spectrum}),
\label{eq:Luminosity_formula}
\end{equation} where $P$ denotes the peak flux during the burst duration. For \f\, bursts, $P$ given in ${\rm erg.cm^{-2}s^{-1},}$ hence

\begin{equation}
k(z)=\dfrac{\int_{1\,{\rm keV}}^{10^{4}\,{\rm keV}}E.S(E)dE}{\int_{(1+z)E_{min}}^{(1+z)E_{max}}E.S(E)dE} \, ;
\label{eq:definition_of_k--Fermi}
\end{equation} where $S(E)$ describes the observed spectrum of the burst, while for \B\, and \s\, bursts with $P$ given in ${\rm ph.cm^{-2}s^{-1}}$,

\begin{equation}
k(z)=\dfrac{\int_{1\,{\rm keV}}^{10^{4}\,{\rm keV}}E.S(E)dE}{\int_{(1+z)E_{min}}^{(1+z)E_{max}}S(E)dE} \, .
\label{eq:definition_of_k---Swift}
\end{equation} 

For all GRBs for which spectral parameter measurements are not available, the spectral energy peak, $E_p$, is randomly sampled from that of the observed distribution of \f\, GRBs with spectral measurements, following \cite{Paul-2018-MNRAS}, which demonstrated that this method statistically reproduces the pseudo-redshifts of all long GRBs with known redshifts. The sample of \f\, bursts are found to have a log-normal distribution of $E_p$, with $ <E_p> = 382.8 $ keV; moreover, $ <\alpha> = -0.2 $ and $ <\beta> = -3.5 $. The justification behind this is as follows: \f -GBM being a wide-band GRB detector, samples the $E_p$ space without any selection bias. This is evident from the fact that the $k(z)$ for \f\, deviates significantly from unity only at very high redshifts (see Fig. 1 of \cite{Paul-2018-MNRAS}), where the formation rate of GRBs is itself extremely low due to the absence of the progenitors. Hence, the spectral parameter distribution of \f -GRBs is representative of the true GRB population. By randomly selecting $E_p$ from the observed distribution of \f\, bursts, the true distribution of $E_p$ of bursts is being sampled, and there is no need to additionally model this distribution. In doing so, no claim as to the accuracy of the individual values of $E_p$ is claimed, and hence neither the individual values of pseudo-redshifts. This approach thus assigns pseudo-redshifts to bursts only in the statistical sense. This limitation is however not binding to this work, since luminosities of the bursts estimated from these pseudo-redshifts are used only as a collective sample in modelling the LF.

The number of sGRBs with known redshift and spectral parameters is only $15$. Hence, to test the hypothesis that the estimated pseudo-redshifts are indeed representative of the whole sample, I compare the cumulative distribution of the pseudo-redshifts thus derived for each of the instruments to that of the measured redshifts of a total of $30$ sGRBs, with or without spectral parameters. The 2-sample KS test rules out the hypothesis that any of the pseudo-redshift distributions are drawn from the known redshift population when the full range of redshifts is considered, as shown in the left panel of Fig. \ref{fig:z-distribution}. However, the number of GRBs with observed redshifts is still quite small to draw negative conclusions from this global comparison. The discrepancy can understood to be due to instrumental selection effects that severely limit the detection of GRBs with high redshifts, primarily via the identification of the host galaxy \citep{Berger-2014-sGRB_review}. \cite{Yonetoku_et_al.-2014-ApJ} pointed out that the pseudo-redshift distributions matches well with the measured ones from their sample, when both are limited to a redshift of $1.0.$ In this work it is found that as many as $25$ of the $30$ GRBs are located within this redshift. Given that the progenitor mass available for the production of sGRBs does not reduce drastically at $z>1.0$ from population synthesis studies (see Fig. \ref{fig:CSFR} and Section \ref{subsec:Modelling the LF}), this is indicative of the fact that selection effects indeed play an important role in the measurement of redshifts of GRBs. When limited to this range, the pseudo-redshift distributions of all the instruments have probabilities $>0.66$ of being drawn from the same population as the known redshift distribution, see right panel of Fig. \ref{fig:z-distribution}. Hence the pseudo-redshifts can be safely used to calculate the luminosities for all the bursts with unknown redshifts. The resultant $L$-$z$ distributions of \f\, and \B\, GRBs are shown in Fig. \ref{fig:L-z}. This approach mitigates the statistical limitation of a sample of redshift-measured bursts, and also the selection bias that plagues the very measurement of redshift. It is to be noted that studies that model the luminosity function considering only the short bursts with measured redshifts, are hence not representative of the true sample.

\subsection{Modelling the luminosity function}
\label{subsec:Modelling the LF}

\begin{figure}
\centering{}
\includegraphics[scale=0.5]{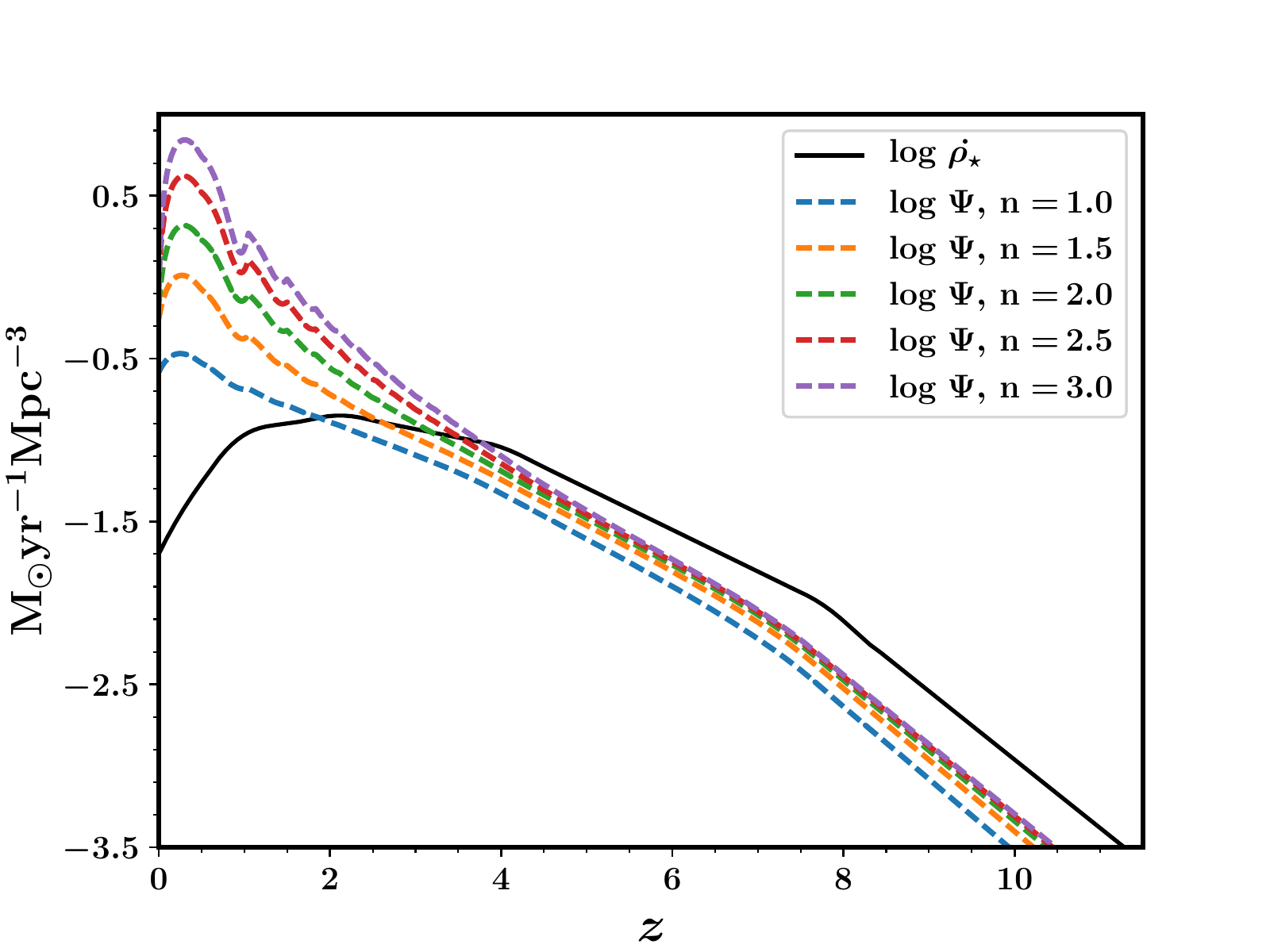}
\caption{The cosmic star formation rate taken from \citet{Bouwens_et_al.-2015-ApJ}, shown in the black solid line, is convolved with a time-delay distribution (see text) via Equation \ref{eq:CSFR_convolution} to derive the binary coalescence rate for various values of the parameter $n.$ (A coloured version of this figure is available in the online journal.)
\label{fig:CSFR}}
\end{figure}

If the sGRB progenitors are produced by coalescences of neutron star (NS) binaries, then assuming that $ \Psi(z) $ is the effective mass available for coalescence per unit time per unit volume, it follows the cosmic star formation rate $ \dot{\rho_{\star}}(z),$ delayed by a time $\tau$ given by $ \tau[z,z'] = t_{\rm{age}}(z)-t_{\rm{age}}(z'),$ where $t_{\rm{age}}(z)$ is the age of the universe calculated in the standard way in $\Lambda$-cold dark matter cosmology. If the delay distribution is given by $P(\tau)$, then $ \Psi(z) $ at redshift $z$ is given by 
\begin{equation}
\Psi(z) = \intop_{z_{\rm{\rm{min}}}(z)}^{\infty} \dot{\rho_{\star}}(z') \, P\left(\tau[z,z']\right) \frac{d\tau}{dz'} dz',
\label{eq:CSFR_convolution}
\end{equation} where $z_{\rm{min}}(z)$ is obtained on solving $ t_{\rm{age}}(z) - t_{\rm{age}}(z_{\rm{min}}) = \tau_{\rm{min}}. $ The probability distribution function $P(\tau)$ is normalized over the chosen range of $\tau,$ bounded below by $\tau_{\rm{min}}$.

$\Psi(z)$ is shown in Fig. \ref{fig:CSFR} by convolving the cosmic star formation rate obtained from \cite{Bouwens_et_al.-2015-ApJ} with a delay distribution of the form $ P(\tau) \propto \tau^{-n}, $ for various values of $n.$ A number of population synthesis codes \citep{Schneider_et_al.-2001-MNRAS, Belczynski_et_al.-2006-ApJ, O'Shaughnessy_et_al.-2008-ApJ} have studied the rate of binary coalescences, concluding that the delay distribution is typically well-approximated as $ P(\tau) \propto \tau^{-1} $ with $ \tau_{\rm{min}} = 10$ Myr. For the rest of the work, I let $n$ vary between $1$, $1.5$ and $2.0$ for the sake of generality. Variation in the choice of $\tau_{\rm{min}}$ in the same order of magnitude has no significant effect on the convolved rate, $\Psi(z).$

\begin{figure}
\centering{}
\includegraphics[scale=0.5]{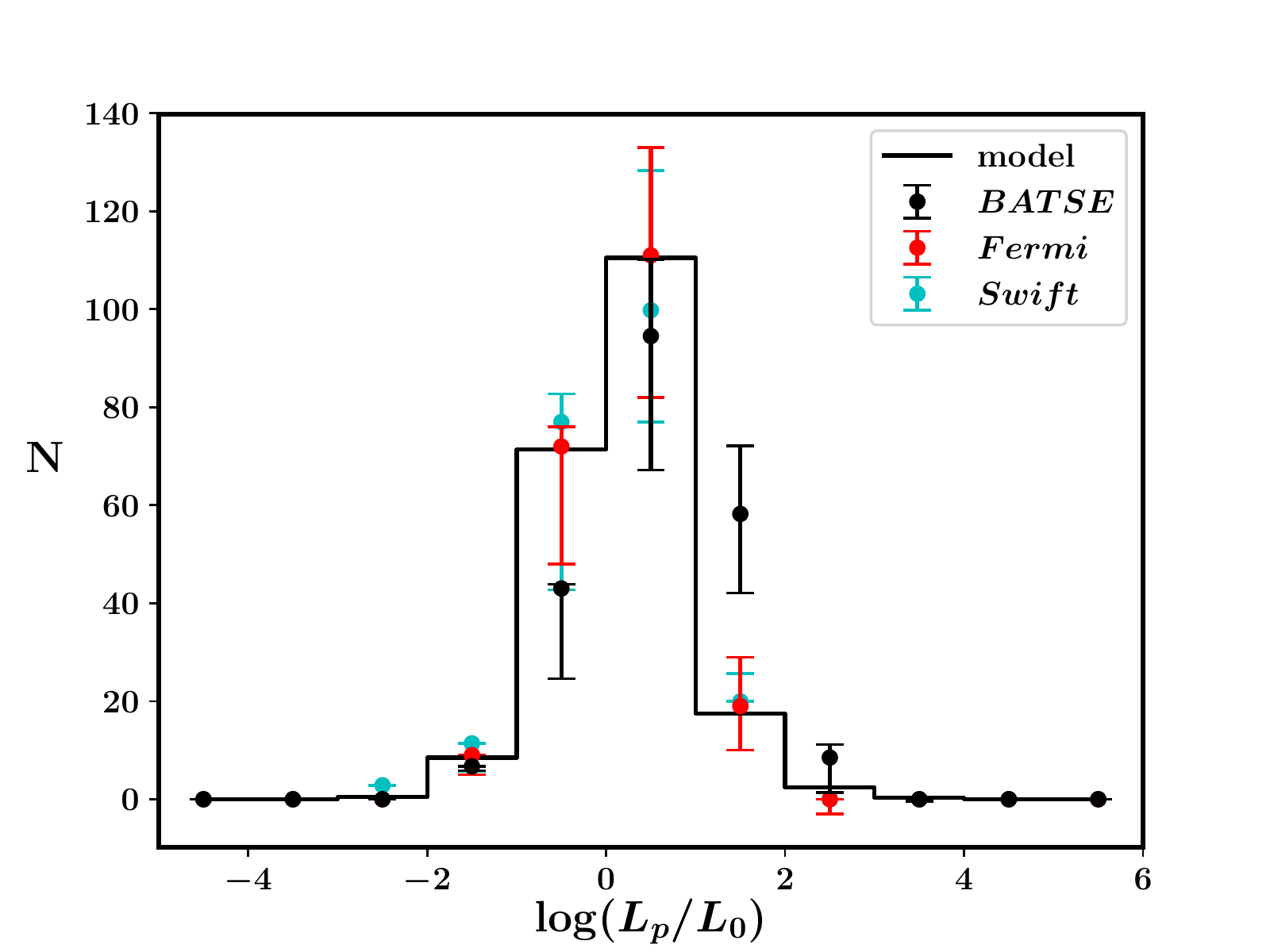}
\caption{The error-bars are the data for \B\, (black), \f\, (red) and \s\, (cyan). It is clearly seen that the \B\, data clearly deviates from the \f\, and \s\, data. The $\Gamma = 0, \; n = 1.0$ BPL model (see Table \ref{tab:BPL_fits}) has been plotted as a thick-line, and the bestfit \B\, model with non-zero $\Gamma$ has not been plotted for simplicity. All the plots are normalized to \f; $L_{0}=10^{52}{\rm \, erg \, s^{-1}}$. (A coloured version of this figure is available in the online journal.)
\label{fig:data-vs-model}}
\end{figure}

Whereas \cite{Yonetoku_et_al.-2014-ApJ} used only $72$ \B\, GRBs with spectral parameters that they estimated, as a result of sampling the $E_p$ from the observed distribution of that of \f, I have a large number of bursts available to model the sGRB LF via the luminosities computed via the pseudo redshifts, as discussed before. This approach allows a range of models to be tested and sufficient confidence be placed on the parameters of the bestfit model. Moreover, the $E_p$ measurements have been directly used for the \f\, GRBs whenever available, whereas \cite{Ghirlanda_et_al.-2016-A&A} has used the Yonetoku correlation to model the \f\, $E_p$-distribution via the LF. In Fig. \ref{fig:data-vs-model} is shown the luminosity distribution of all the GRBs from the three instruments (along with a bestfit model, see below); all the curves are normalized to that of \f. The total number of GRBs used for the different instruments are tabulated in Table \ref{tab:GRBs_used_for_modelling_LF}.

The model fits are carried out via the standard Levenberg-Marquardt algorithm of minimizing the discrepancy defined as $ d^2 = (model - data)^{2} $, available in the Python library \textbf{scipy}\footnote{\url{https://docs.scipy.org/doc/scipy/reference/generated/scipy.optimize.curve_fit.html}}.

Attempts are first made to fit a simple powerlaw (SPL) model of the LF, $\Phi_z(L) \propto L^{-\nu}$, with $ \nu \in [0.01,10.0].$ Table \ref{tab:SPL_fits} lists the reduced chisquared, $\red$, for $10$ degrees of freedom, for the chosen values of $n$. It is clearly seen that this model is ruled out for all three instruments for the whole range of $\nu$ with a high degree of confidence. This rules out the conclusion of \cite{Yonetoku_et_al.-2014-ApJ}, who found the LF to be well-described by a simple powerlaw of index $1,$ while supporting and extending the conclusion of \cite{Ghirlanda_et_al.-2016-A&A}, who ruled out this model with $ \nu > 2.0. $ The large number of GRBs in the present dataset helps in reaching this conclusion.

\begin{table}
\caption{The best fits to the SPL model and the corresponding reduced chisquared ($\red$), corresponding to $10$ degrees of freedom.
\label{tab:SPL_fits}}
\begin{center}
\begin{tabular}{|c|c|c|c|c|}
\hline
$n$ & parameters & \B & \f & \s \tabularnewline
\hline
\multirow{2}{*}{$1.0$} & $\nu$ & $1.12$ & $1.23$ & $1.37$ \tabularnewline
 & $\red$ & $233.1$ & $26.5$ & $10.1$ \tabularnewline
\hline
\multirow{2}{*}{$1.5$} & $\nu$ & $1.10$ & $1.20$ & $1.33$ \tabularnewline
 & $\red$ & $276.5$ & $35.4$ & $10.9$ \tabularnewline
\hline
\multirow{2}{*}{$2.0$} & $\nu$ & $1.09$ & $1.18$ & $1.31$ \tabularnewline
 & $\red$ & $300.6$ & $39.4$ & $11.2$ \tabularnewline
\hline
\end{tabular}
\end{center}
\end{table}

\begin{table}
\caption{The best fits to the ECPL model and the corresponding reduced chisquared ($\red$), corresponding to $8$ degrees of freedom. Here, $L_{\rm{b}}$ is given in units of $L_{0} = 10^{52}{\rm \, erg \, s^{-1}}$. The best fits to $\nu$ and $L_{\rm{b}}$ are not provided for the three instruments separately since the same values are applicable to all instruments; whereas $\Gamma$ and $\red$ vary with instruments for the same values for $\nu$ and $L_{\rm{b}}$. The errors refer to $1$-$\sigma$ uncertainties.
\label{tab:ECPL_fits}}
\begin{center}
\begin{tabular}{|c|c|c|c|c|c|}
\hline
$n$ & parameters &  & \f & \s & \B \tabularnewline
\hline
\multirow{4}{*}{$1.0$} & $\nu$ & $0.71 ^{+0.05} _{-0.36}$ & & & \tabularnewline
 & $L_{\rm{b}}$ & $7.42 ^{+7.21} _{-1.96}$ & & & \tabularnewline
 & $\Gamma$ & & $0.00$ & $0.00$ & $0.41 ^{+0.15} _{-0.12}$ \tabularnewline
 & $\red$ & & $0.31$ & $0.21$ & $0.75$ \tabularnewline
\hline
\multirow{4}{*}{$1.5$} & $\nu$ & $0.64 ^{+0.05} _{-0.39}$ & & & \tabularnewline
 & $L_{\rm{b}}$ & $6.84 ^{+6.73} _{-1.58}$ & & & \tabularnewline 
 & $\Gamma$ & & $0.00$ & $0.00$ & $0.38 ^{+0.13} _{-0.10}$ \tabularnewline
 & $\red$ & & $0.39$ & $0.19$ & $0.82$ \tabularnewline
\hline
\multirow{4}{*}{$2.0$} & $\nu$ & $0.60 ^{+0.05} _{-0.38}$ & & & \tabularnewline
 & $L_{\rm{b}}$ & $6.61 ^{+6.09} _{-1.53}$ & & & \tabularnewline
 & $\Gamma$ & & $0.00$ & $0.00$ & $0.36 ^{+0.12} _{-0.09}$ \tabularnewline
 & $\red$ & & $0.41$ & $0.19$ & $0.84$ \tabularnewline
\hline
\end{tabular}
\end{center}
\end{table}

\begin{table}
\caption{The best fits to the BPL model and the corresponding reduced chisquared ($\red$), corresponding to $7$ degrees of freedom. Here, $L_{\rm{b}}$ is given in units of $L_{0} = 10^{52}{\rm \, erg \, s^{-1}}$. The best fits to $\nu_{1}$, $\nu_{2}$ and $L_{\rm{b}}$ are not provided for the three instruments separately since the same values are applicable to all instruments; whereas $\Gamma$ and $\red$ vary with instruments for the same values for $\nu_{1}$, $\nu_{2}$ and $L_{\rm{b}}$. The errors refer to $1$-$\sigma$ uncertainties.
\label{tab:BPL_fits}}
\begin{center}
\begin{tabular}{|c|c|c|c|c|c|}
\hline
$n$ & parameters &  & \f & \s & \B \tabularnewline
\hline
\multirow{5}{*}{$1.0$} & $\nu_{1}$ & $0.48 ^{+0.22} _{-0.48}$ & & & \tabularnewline
 & $\nu_{2}$ & $1.86 ^{+1.08} _{-0.20}$ & & & \tabularnewline
 & $L_{\rm{b}}$ & $1.52 ^{+1.58} _{-0.67}$ & & & \tabularnewline
 & $\Gamma$ & & $0.00$ & $0.00$ & $0.17 ^{+0.05} _{-0.05}$ \tabularnewline
 & $\red$ & & $0.10$ & $0.42$ & $1.09$ \tabularnewline
\hline
\multirow{5}{*}{$1.5$} & $\nu_{1}$ & $0.38 ^{+0.23} _{-0.38}$ & & & \tabularnewline
 & $\nu_{2}$ & $1.85 ^{+1.04} _{-0.19}$ & & & \tabularnewline
 & $L_{\rm{b}}$ & $1.46 ^{+1.36} _{-0.62}$ & & & \tabularnewline 
 & $\Gamma$ & & $0.00$ & $0.00$ & $0.16 ^{+0.04} _{-0.05}$ \tabularnewline
 & $\red$ & & $0.10$ & $0.39$ & $1.09$ \tabularnewline
\hline
\multirow{5}{*}{$2.0$} & $\nu_{1}$ & $0.34 ^{+0.23} _{-0.34}$ & & & \tabularnewline
 & $\nu_{2}$ & $1.85 ^{+1.03} _{-0.19}$ & & & \tabularnewline
 & $L_{\rm{b}}$ & $1.45 ^{+1.32} _{-0.60}$ & & & \tabularnewline
 & $\Gamma$ & & $0.00$ & $0.00$ & $0.15 ^{+0.04} _{-0.05}$ \tabularnewline
 & $\red$ & & $0.10$ & $0.39$ & $1.09$ \tabularnewline
\hline
\end{tabular}
\end{center}
\end{table}

Next, the observed distributions are fit to the exponential cutoff powerlaw (ECPL) model:

\begin{equation}
\Phi_z(L) = \Phi_{0}
\left(\frac{L}{L_{b}}\right)^{-\nu} \exp\left[- \left(\frac{L}{L_{b}}\right) \right],
\label{eq:The-ECPL-model}
\end{equation} and the broken powerlaw (BPL) model:

\begin{equation}
\Phi_z(L)=\Phi_{0}\begin{cases}
\left(\frac{L}{L_{b}}\right)^{-\nu_{1}}, & L\leq L_{b}\\
\left(\frac{L}{L_{b}}\right)^{-\nu_{2}}, & L>L_{b}.
\end{cases}
\label{eq:The-BPL-model}
\end{equation}

It is seen that, although the \f\, and the \s\, data are well fit by both the models, the \B\, data (see Fig. \ref{fig:data-vs-model}) is not. This is clearly understood to be due to the fact that the \B\, data is significantly different from the \f\, and the \s\, data, specially at higher luminosities. While writing Equation \ref{eq:definition_of_phi}, it was assumed that the probability of detection of a burst is $0$ below $L_c$ and $1$ above it. However, that may not be the case for all instruments, the change being gradual. This effect can be modelled by introducing a detection probability that is a function of the observed flux $P$, given as $D(P)$, thus modifying Equation \ref{eq:definition_of_phi} to

\begin{equation}
N(L_{1},L_{2};z_{1},z_{2}) = T \, \dfrac{\Delta\Omega}{4\pi} \,
\intop_{z_{1}}^{z_{2}} dV
\intop_{\rm{max}[L_1,\, L_c]}^{L_2} dL \; \Phi_z(L) \dfrac{\Rdot(z)}{1+z} \times D(L,z),
\label{eq:introduction_of_D}
\end{equation} where $D(L,z) \equiv D(P)$. Assuming
\begin{equation}
D(P) \propto P^{\Gamma},
\label{eq:form_of_D}
\end{equation} the normalization is defined such that $ \Phi_{z}(L) \, D(L,z) $ is normalized in the absolute limits. The data is then fit to the model keeping $\Gamma$ as a free parameter for each instrument. It is envisaged that the same set of parameters for $\Phi_{z}(L)$ describes the data of each instrument, whereas $\Gamma$ itself may be different for the different instruments. That is indeed the case, with $\Gamma$ being consistent with $0$ for both \f\, and \s, whereas non-zero for the \B\, data.

The combined bestfits to the ECPL and BPL models are given in Table \ref{tab:ECPL_fits} and \ref{tab:BPL_fits} respectively. It is seen that the low luminosity index ($\nu_1$) of the BPL model is weakly constrained from below, although the other parameters are well constrained. It is noted that the the BPL model fits are consistent with the $68 \%$ confidence intervals quoted for this model by \cite{Ghirlanda_et_al.-2016-A&A}, for all three scenarios considered by them. From the current dataset, it is impossible to distinguish between the ECPL and BPL bestfit models, as the relative errors on the luminosity are large due to large propagated errors on the estimated luminosities ($40 \%$ on an average). Although the break luminosity ($L_{\rm{b}}$) is weakly constrained from above for the ECPL model, the robust lower limits makes it a few times larger as compared to the BPL model, same as in long GRBs \citep{Amaral-Rogers_et_al.-2017-MNRAS, Paul-2018-MNRAS}.

\subsection{The local GRB rate}
\label{subsec:The local GRB rate}

\begin{table}
\caption{The bestfit normalizations for the models. The errors refer to $1$-$\sigma$ uncertainties obtained on propagating the errors in the fitted parameters quoted in Tables \ref{tab:ECPL_fits} and \ref{tab:BPL_fits}. The range of the local GRB formation rate uncorrected for the beaming factor, $\Rdot(0)$, refers to $68 \%$ confidence limits combining the two models.
\label{tab:fBC0_fits}}
\begin{center}
\begin{tabular}{|c|c|c|c|}
\hline 
$n$ & model & $f_{\rm{B}}C(0)$ & $\Rdot(0)$ \tabularnewline
 & & $[10^{-9} \,{\rm M_{\odot}^{-1}}]$ & $ [\rm{ yr^{-1} Gpc^{-3} }] $ \tabularnewline
\hline
\multirow{2}{*}{$1.0$} & ECPL & $13.7 ^{+1.2} _{-3.9}$ & \multirow{2}{*}{$0.68$-$3.89$} \tabularnewline
 & BPL & $3.74 ^{+3.76} _{-1.15}$ \tabularnewline
\hline
\multirow{2}{*}{$1.5$} & ECPL & $6.45 ^{+0.39} _{-1.32}$ & \multirow{2}{*}{$0.82$-$3.80$} \tabularnewline
 & BPL & $2.05 ^{+1.73} _{-0.58}$ \tabularnewline
\hline
\multirow{2}{*}{$2.0$} & ECPL & $3.65 ^{+0.26} _{-0.61}$ & \multirow{2}{*}{$0.61$-$2.66$} \tabularnewline
 & BPL & $1.23 ^{+0.94} _{-0.34}$ \tabularnewline
\hline
\end{tabular}
\end{center}
\end{table}

\begin{table}
\caption{Comparison of the derived local GRB formation rate uncorrected for the beaming factor, $\Rdot(0)$ , with previous works. The rate quoted for the present work combines the results of all considered $n$-s and includes the $68 \%$ confidence intervals of both the models.
\label{tab:rate_comparison}}
\begin{center}
\begin{tabular}{|c|c|}
\hline 
Reference & $\Rdot(0)$ \tabularnewline
 & $ [\rm{ yr^{-1} Gpc^{-3} }] $ \tabularnewline
\hline
\hline
Ghirlanda et al. (2016), model [a] & $0.13$-$0.24$ \tabularnewline
\hline
Guetta \& Piran (2005) & $0.1$-$0.8$ \tabularnewline
Yonetoku et al. (2014) & $0.24$-$0.94$ \tabularnewline
Ghirlanda et al. (2016), model [c] & $0.65$-$1.10$ \tabularnewline
present work & $0.61$-$3.89$ \tabularnewline
\hline
Coward et al. (2012) & $5$-$13$ \tabularnewline
Guetta \& Piran (2006) & $8$-$30$ \tabularnewline
\hline
\end{tabular}
\end{center}
\end{table}

\begin{figure*}
\centering{}
\includegraphics[scale=0.6]{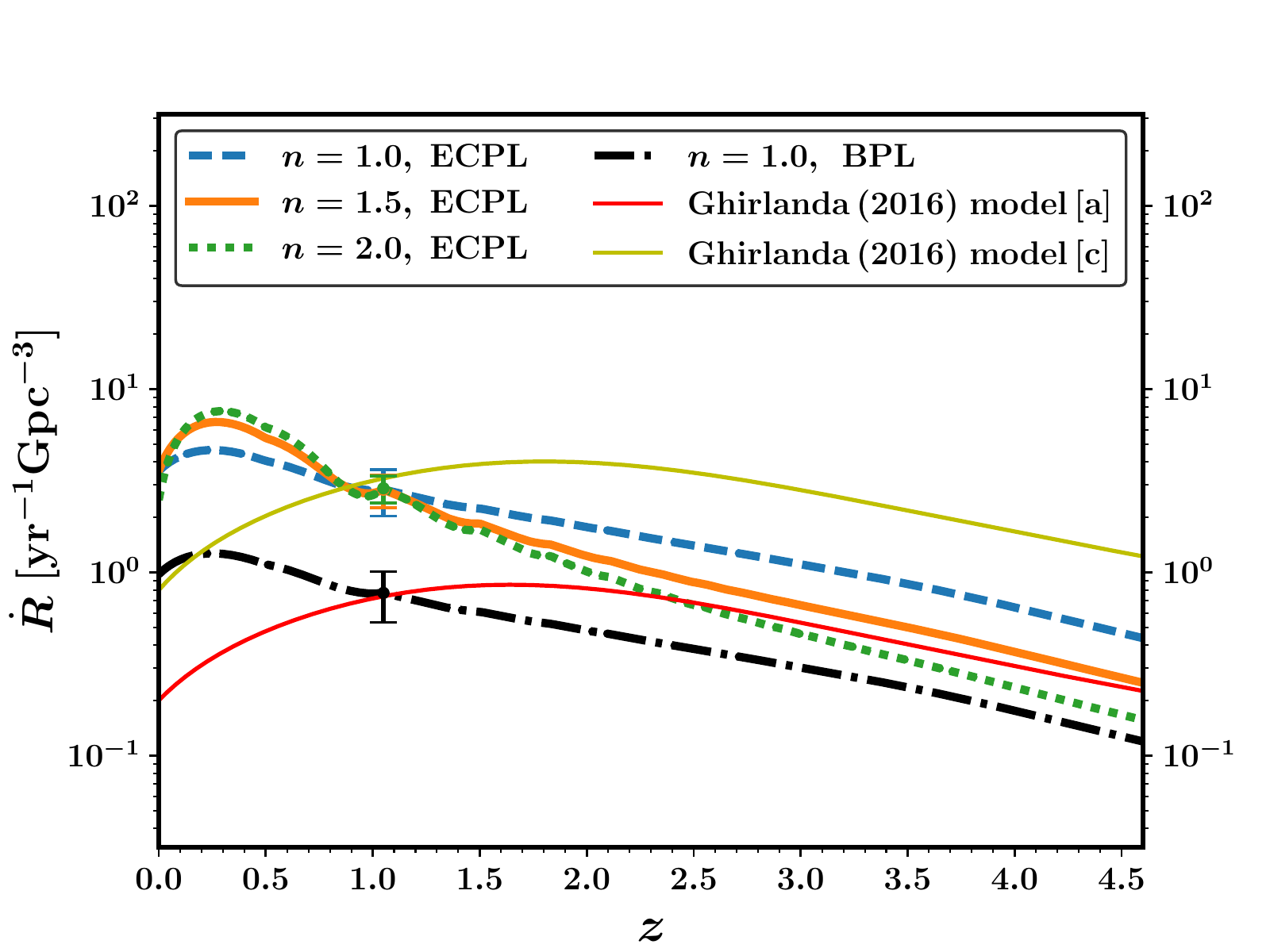}
\caption{The beaming-uncorrected short GRB rate as a function of the redshift, $ \Rdot(z) $. The (blue) dashed, (orange) solid and (green) dotted lines correspond to the results from the ECPl model for $n=1.0$, $1.5$ and $2.0$ respectively. Typical $1$-$\sigma$ errorbars have been shown at a representative redshift $\sim 1$. It is seen that the dependence on $n$ is rather weak. The (black) dot-dashed line represents the $n=1.0$ BPL model. $ \Rdot(0)$ is a few times smaller for the BPL model compared to the ECPL model for each $n$. Since the general dependence with redshift is similar for the two models, only one curve has been shown for the BPL model for simplicity. The red and yellow thick lines represent the model fits from \citet{Ghirlanda_et_al.-2016-A&A} (see their Fig. 4). Their $ \Rdot(0)$ are smaller compared to the present work (see Table \ref{tab:rate_comparison}). However, their model [a] curve converges with those of the present work at higher redshifts, although model [c] is a factor of few higher than the present curves at high redshifts. (A coloured version of this figure is available in the online journal.)
\label{fig:Rdot_of_z}}
\end{figure*}

The normalization of the models are kept free during the fits, and can thus be derived via the solutions. With the knowledge of $T \sim 8.9$ yr and assuming $\frac{\Delta\Omega}{4\pi} \sim \frac{1}{3}$ for \f, the ratios of the observed and modelled normalizations (for the corresponding models in Section \ref{subsec:Modelling the LF}) are converted to derive $f_{\rm{B}}C(0)$, which are used to derive the detected sGRB rate via $ \Rdot(z) = f_{\rm{B}}C(0) \, \Phi(z) $. These, along with the propagated errors, are listed in Table \ref{tab:fBC0_fits}, along with the combined $68\%$ confidence intervals of $\Rdot(0)$ combining both the ECPL and BPL models. $\Rdot(z)$ is plotted in Fig. \ref{fig:Rdot_of_z}, along with those derived by \cite{Ghirlanda_et_al.-2016-A&A}. It is seen that this quantity depends weakly on $n$.

It is seen that the deduced GRB detection rate is very weakly dependent on the delay-distribution. Combining the results, one gets $ \Rdot(0) \sim 0.61 $-$3.89 \, \rm{ yr^{-1} Gpc^{-3} } $. While clearly higher than model (a) of \cite{Ghirlanda_et_al.-2016-A&A}, this is consistent with the higher end of \cite{Guetta_and_Piran-2005-A&A}, \cite{Yonetoku_et_al.-2014-ApJ} and model (c) of \cite{Ghirlanda_et_al.-2016-A&A}, while being smaller than the rates deduced by \cite{Guetta_and_Piran-2006-A&A} and \cite{Coward_et_al.-2012-MNRAS}. This comparison is summarized in Table \ref{tab:rate_comparison}.

\begin{figure*}
\centering{}
\includegraphics[scale=0.6]{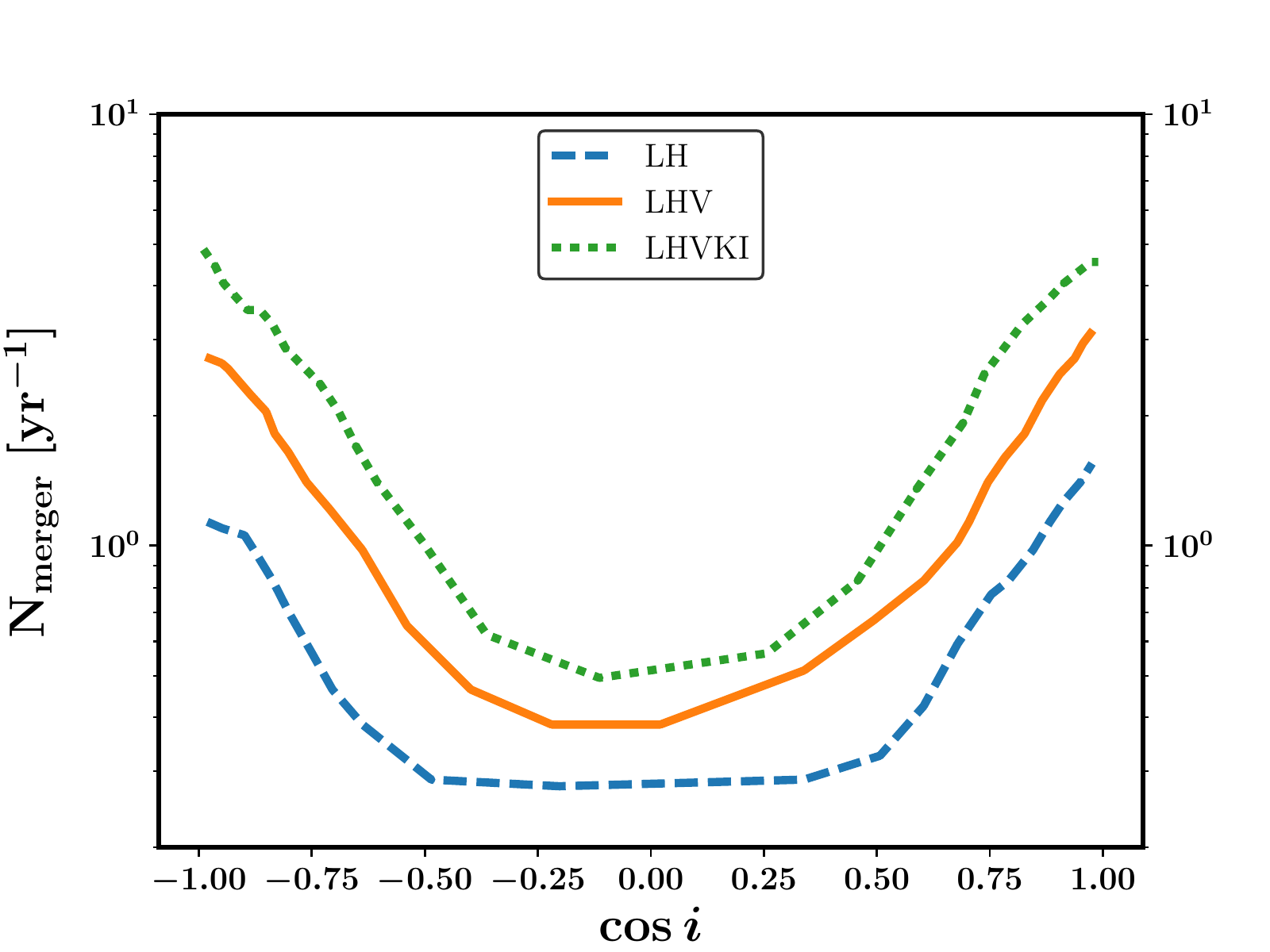}
\caption{The lower limit of the binary neutron star merger rate (BNSM) as a function of the inclination of the normal to the merger plane with respect to the line of sight of the observer, $i$, for the LH (Livingston-Hanford) configuration in (blue) dashed line, the LHV (Livingston-Hanford-Virgo) configuration in (orange) solid line, and the future LHVKI (Livingston-Hanford-Virgo-KAGRA-India) configuration in (green) dotted line. Owing to the weak dependence of the deduced $ \Rdot(z) $ on $n$ (see Fig. \ref{fig:Rdot_of_z}), the most likely scenario (from population synthesis studies) of $n=1.0$ has been shown. The data for the limiting distance of the GW networks used for this purpose have been taken from \citet{Saleem_et_al.-2018-MNRAS}. As expected, the detection rate is larger for face-on systems ($\cos i \to \pm 1$) than edge-on systems ($\cos i \sim 0$). (A coloured version of this figure is available in the online journal.)
\label{fig:merger-rate}}
\end{figure*}

\subsection{Prediction for CZTI}
\label{subsec:CZTI predictions}

Combining the model parameters and the derived normalizations, predictions are made for the rate of sGRBs detectable by the \AS\, \citep{Rao_et_al.-2016-arXiv-Astrosat} hard X-ray detector CZTI \citep{Rao_et_al.-2016-ApJ, Bhalerao_et_al.-2017-JApA}, similar to \cite{Paul-2018-MNRAS} who predicted a sizeable under-detection for long GRBs. Assuming $\frac{\Delta\Omega}{4\pi} \sim \frac{1}{3}$, the combined model rate comes out in the range of $14$-$42$ per year at $68 \%$ confidence. However, in the last two years of operation, it has detected only $11$ sGRBs by triggered searches, i.e. by subjective search of GRBs from automatic triggers by other satellites. Moreover, the searches have been carried out at coarse time-bins due to the uncertainties in the characterization of noise at finer time bins. This study implies that a careful automatic search of the CZTI data post removal of sub-second noise in the data will reveal at least $\sim 20$ sGRBs hidden till date. A careful analysis of the sub-second noise is being carried out and will be reported elsewhere. The importance of an automatic detection algorithm and alerts to the astronomical community for quick follow-up measurements cannot be underestimated.

\section{The Binary Coalescence Rate}
\label{sec:Event rate}

The observed event rate of sGRBs can be corrected for the beaming factor $ f_{\rm{B}} = 1 - \cos(\theta_j), $ where $ \theta_j $ is the half-opening angle of the jet, to derive their true sky rate:
\begin{equation}
R_0 = \dfrac{ \Rdot(0) }{ f_{\rm{B}} } .
\end{equation} Using radio to X-ray afterglow observations of 11 bursts upto 2015, \cite{Fong_et_al.-2015-ApJ} constrained the range of $ \theta_j$ to $6$-$26^{\circ}.$ Allowing for the lower limit of the range to be $3^{\circ}$ as derived for GRB111002A \citep{Fong_et_al.-2012-ApJ} and GRB111117A \citep{Margutti_et_al.-2012-ApJ}, the conservative range of $3$-$26^{\circ}$ is used along with $ \Rdot(0) $ deduced in the previous section to derive $R_0$. The $68\%$ confidence ranges are given by $6.72$-$2838 \, \rm{ yr^{-1} Gpc^{-3} } $ for $n=1.0$; $8.10$-$2773 \, \rm{ yr^{-1} Gpc^{-3} } $ for $n=1.5$; and $6.03$-$1941 \, \rm{ yr^{-1} Gpc^{-3} } $ for $n=2.0$. It is to be noted that although the upper limit of $R_0$ is sensitive to the lower limit of $\theta_j$ and hence debatable, the lower limit of $R_0$ depends on the upper limit of $\theta_j$ and is hence fairly robust. Thus, a sharp lower limit of $ R_0 \sim 6 \, \rm{ yr^{-1} Gpc^{-3} } $ is placed via this work upto $68 \%$ confidence. Assuming that each BNSM produces a sGRB, this is also the minimum rate of BNSMs; if not, the merger rate is higher.

\cite{Saleem_et_al.-2018-MNRAS} has simulated a large sample of mergers of non-spinning NSs with component masses of $1.4 \rm{M_{\odot}}$ each, and taking into account the antenna pattern functions of the gravitational wave detectors, calculated the signal-to-noise ratio (SNR) for their detection as a function of the distance to the merger, and inclination of the axis of the merger plane to the line of sight of the observer, $i$. With a detection criterion set to SNR $>8.0$, this produces the limiting distance ($D_L$) versus inclination ($i$) scatter plot for a combination of detectors: (a) the LH network comprising the Livingston and Hanford detectors, (b) the LHV network with the addition of the Virgo detector, and (c) the LHVKI network, including the KAGRA detector under construction in Japan \citep{Aso_et_al.-2013-PhRvD}, and the approved LIGO-India detector\footnote{ \url{https://dcc.ligo.org/LIGO-M1100296/public} } which is expected to come up in the next decade  (see their Fig. 1 for configurations b and c). In this work, I have used this simulated dataset and integrated the $68 \%$ lower limit of $ \Rdot(z)$ obtained in this work, upto the limiting redshift corresponding to $D_L$, to obtain the total rate as a function of $i$. Since the lower limits are very weakly dependent on $n$, the curves obtained for $n=1.0$, the most likely scenario from population synthesis studies, is shown in Fig. \ref{fig:merger-rate}. Giving equal weights to all $i$, the integrated rates are $ 0.95 \pyr $ for the LH network, $ 1.87 \pyr $ for the LHV network and $ 3.11 \pyr $ for the LHVKI network.

In the few years of the observing run of the LH and the LHV network, there have been five confirmed detections of black hole binary mergers, that of GW150914 \citep{GW150914-2016}, GW151226 \citep{GW151226-2016}, GW170104 \citep{GW170104-2017}, GW170608 \citep{GW170608-2017}, and GW170814 \citep{GW170814-2017}; and one confirmed detection of neutron star binary merger, GW170817 \citep{GW170817-2017}. The derived minimum integrated rate of $ 0.95 \pyr $ for the detection of BNSMs by the LH network is consistent with the detection of the single neutron star inspiral GW170817 that was extensively followed up across the electromagnetic spectrum (EM170817; \cite{EM170817-2017}). In the future runs, the number is expected to increase by a factor of few, see Fig. \ref{fig:merger-rate}.

On the basis of the gravitational wave (GW) data alone from GW170817, \cite{GW170817-2017} placed the rate of BNSMs at $320$-$4740 \, \rm{ yr^{-1} Gpc^{-3} } $ at $90\%$ confidence. This rate is consistent but significantly higher than the sGRB rate derived in this work, $ R_0 \sim 6 $-$ 2838 \, \rm{ yr^{-1} Gpc^{-3} } $. This implies that the fraction of GRBs produced from the BNSMs, $f_{GRB}$, may be smaller than unity. This has important implications in the physics of the mergers, implying that a fraction of the mergers may not be able to produce the classic on-axis jet that are hypothesized to cause sGRBs associated with the gravitational waves \citep{Narayan_et_al.-1992-ApJ}. The on-axis jet scenario has indeed been ruled out for GW/EM170817 by \cite{Kasliwal_et_al.-2017-Science}, who proposed a cocoon model to explain the multi-wavelength electromagnetic observations. Given the large uncertainties from both the GW and the sGRB rates, only a very weak lower limit of $ f_{GRB} > 0.001 $ can be obtained. As the upcoming runs of the GW networks will significantly improve the detection of the BNSMs, it is envisaged that similar extensive follow-up campaigns of the electromagnetic counterparts of these mergers will shed more light on the physical processes surrounding the merger and the evolution of the associated ejecta.

\section{Conclusions}
\label{sec:Conclusions}

In this work, I have combined the accurate spectral energy and redshift measurements of 15 sGRBS available till date, and found a significant linear correlation between the spectral energy peak in the source frame with the source luminosity, also known as the `Yonetoku correlation'. Next I have used this correlation to derive `pseudo-redshifts' of all sGRBs with measured flux, including \B, \s, and \f\, GRBs. Although the redshift distributions of the sample of $30$ sGRBs with known redshifts are not reproduced for the full redshift range, it is found that $25$ of these GRBs are located at $z<1.0$, against the expectations from population synthesis studies. Furthermore, the pseudo-redshifts of all instruments agree well with the observed redshift distribution when limited to this redshift range. Thus, instrumental selection effects are understood to play a role in the non-detection of higher-redshift sGRBs. This provided confidence to use the pseudo-redshifts of the full catalogues to calculate their luminosities. This method does not claim to accurately predict the redshifts of individual bursts, but successfully mitigates the problems of having a statistically limited as well as selectively biased sample of bursts for the study of the luminosity function.

Assuming standard delay between the cosmic star formation and the binary neutron star mergers, which are thought to the progenitors of sGRBs, I attempted to fit the observed luminosity distribution of the largest sGRB sample of $757$ bursts studied till date. The simple powerlaw model of the LF is ruled out with high confidence. Both the exponential cutoff powerlaw (ECPL) and the broken powerlaw (BPL) model are found to fit the data of all three GRB-detectors, with the additional complication that the detection probability is different for \B\, compared to \f\, and \s. It is not possible from the current dataset to compare between the quality of fits between these two models, however. The low-luminosity index of the BPL model ($\nu_1$) is found to be weakly constrained below, although the constraints on the higher luminosity index $\nu_2 \sim 1.85$ and the break luminosity $ L_{\rm{b}} \sim 1.50$ are much tighter. For the ECPL model, the powerlaw index $ \nu \sim 0.7 $ is well-constrained, and although the break luminosity is weakly constrained above, it is at least a few times higher than for the BPL model. Unlike in the case of long GRBs, it is not necessary to invoke any redshift dependence of the break luminosity, consistent with existing works in the sGRB literature. The current work is purely empirical in nature, and does not attempt to provide physical explanation of the LF models, which should be independently pursued via detailed phenomenological models of sGRBs.

The bestfit models are then used to make predictions of the sGRB detection rate of \AS -CZTI, implying that at least $\sim 20$ GRBs are undiscovered till date in the CZTI data by subjective triggered searches. The models are also used to derive the observed event rate of sGRBs, which is found to be weakly dependent on the assumed delay distribution. Adopting conservative limits of the jet opening angle, this is converted to get the true event rate of sGRBs. Assuming that each sGRB is produced from a binary neutron star merger (BNSM), this rate is then used to calculate the rate of BNSMs detectable by the past, current and upcoming global GW detector networks. Robust lower limits of $1.87 \pyr$ for the LHV and $3.11 \pyr$ for the LHVKI networks are derived, while the true rates may be significantly higher. The uncertainty on the rate of BNSMs calculated via the only confirmed BNSM detection via gravitational waves presented in the discovery paper of GW170817 \citep{GW170817-2017}, as well as the uncertainty on the sGRB rate derived from this work, are large. This makes it impossible to rule out the scenario that not all mergers produce sGRBs. However, the presence of a tension between these independently derived rates can have significant implications on the physics of the merger ejecta, in line with the detailed study of the electromagnetic follow-up of GW170817 \citep{Kasliwal_et_al.-2017-Science}. Similar extensive electromagnetic follow-up campaigns of the future BNSMs detected via gravitational waves will be able to make more conclusive statements about the physics of the merger ejecta on a case-by-case basis.

\section*{Acknowledgements}
I extend my sincere thanks to the anonymous referee for their critical comments, which significantly improved the quality of both the work and the manuscript; my Ph.D. advisor A.R. Rao for providing the motivation for the work; Patrick Dasgupta for discussions on GRBs throughout the course of the work; \AS -CZTI member Vidushi Sharma for the updated list of GRBs detected by CZTI and related discussions; Marek J. Szczepanczyk, Varun Bhalerao and Shreya Anand for discussions on the aLIGO/VIRGO sensitivities; and Muhammad Saleem for providing the gravitational wave detectors' sensitivity data and related discussions.

\bibliographystyle{apj}
\bibliography{lf}

\end{document}